\documentstyle[11pt,amssymb]{article}

\makeatletter
\@addtoreset{equation}{section}
\makeatother

\def\dalemb#1#2{{\vbox{\hrule height .#2pt
        \hbox{\vrule width.#2pt height#1pt \kern#1pt
                \vrule width.#2pt}
        \hrule height.#2pt}}}
\def\square{\mathord{\dalemb{6.8}{7}\hbox{\hskip1pt}}}

\def\0{{\sst{(0)}}}
\def\1{{\sst{(1)}}}
\def\2{{\sst{(2)}}}
\def\3{{\sst{(3)}}}
\def\4{{\sst{(4)}}}
\def\5{{\sst{(5)}}}
\def\6{{\sst{(6)}}}
\def\7{{\sst{(7)}}}
\def\8{{\sst{(8)}}}
\def\n{{\sst{(n)}}}

\def\R{\rlap{\rm I}\mkern3mu{\rm R}}

\def\ep{\epsilon}
\def\td{\tilde}
\def\wtd{\widetilde}

\def\rr{{\sst{\rm RR}}}
\def\ns{{\sst{\rm NS}}}
\def\enhancon{{enhan\c{c}on}}

\let\a=\alpha \let\b=\beta

\def\nn{\nonumber} \def\bd{\begin{document}} \def\ed{\end{document}}
\def\ds{\documentstyle} \let\fr=\frac \let\bl=\bigl \let\br=\bigr
\let\Br=\Bigr \let\Bl=\Bigl
\let\bm=\bibitem
\let\na=\nabla
\let\pa=\partial \let\ov=\overline
\newcommand{\be}{\begin{equation}}
\newcommand{\ee}{\end{equation}}
\def\ba{\begin{array}}
\def\ea{\end{array}}
\def\ft#1#2{{\textstyle{{\scriptstyle #1}\over {\scriptstyle #2}}}}
\def\fft#1#2{{#1 \over #2}}
\def\del{\partial}
\def\sst#1{{\scriptscriptstyle #1}}
\def\oneone{\rlap 1\mkern4mu{\rm l}}
\def\ie{{\it i.e.\ }}
\def\via{{\it via}}
\def\semi{{\ltimes}}
\def\str{{\rm str}}
\def\jm{{\rm j}}
\def\im{{\rm i}}
\def\bOmega{{{\bar\Omega}}}
\def\Qn{{{Q_{\sst{\rm N}}}}}
\def\tX{{{\wtd X}}}

\def\mapright#1{\smash{\mathop{-\!\!\!-\!\!\!-\!\!\!-\!\!\!-\!\!\!
             \longrightarrow}\limits^{#1}}}
\def\maprightt#1#2{\smash{\mathop{-\!\!\!-\!\!\!-\!\!\!-\!\!\!-\!\!\!
             \longrightarrow}\limits^{#1}_{#2}}}

\newcommand{\ho}[1]{$\, ^{#1}$}
\newcommand{\hoch}[1]{$\, ^{#1}$}
\newcommand{\bea}{\begin{eqnarray}}
\newcommand{\eea}{\end{eqnarray}}
\newcommand{\ra}{\rightarrow}
\newcommand{\lra}{\longrightarrow}
\newcommand{\Lra}{\Leftrightarrow}
\newcommand{\ap}{\alpha^\prime}
\newcommand{\bp}{\tilde \beta^\prime}
\newcommand{\tr}{{\rm tr} }
\newcommand{\Tr}{{\rm Tr} }

\newcommand{\NP}{Nucl. Phys. }
\newcommand{\tamphys}{\it Center for Theoretical Physics\\
Texas A\&M University, College Station, TX 77843}
\newcommand{\umich}{\it Department of Physics\\
University of Michigan, Ann Arbor, Michigan 48109}
\newcommand{\upenn}{\it Department of Physics and Astronomy\\
University of Pennsylvania, Philadelphia,  PA 19104}
\newcommand{\SISSA}{\it  SISSA-ISAS and INFN, Sezione di Trieste\\
Via Beirut 2-4, I-34013, Trieste, Italy}

\newcommand{\ihp}{\it Institut Henri Poincar\'e\\
  11 rue Pierre et Marie Curie, F 75231 Paris Cedex 05}

\newcommand{\auth}{M. Cveti\v{c}\hoch{\dagger1},
H. L\"u\hoch{\star2} and C.N. Pope\hoch{\ddagger3}}

\begin{document}
\begin{flushright}
\hfill{CTP TAMU-32/00}\ \ \ {UPR-910-T}\ \ \
{UM-TH-00-29}\ \ \ {RUNHETC-2000-43}\ \ \
{hep-th/0011023}\\
\hfill{November, 2000}\\
\end{flushright}


\begin{center}
{ \large {\bf Brane Resolution Through Transgression}}

\vspace{10pt}
\auth

\vspace{7pt}
{\hoch{\dagger}\upenn}

\vspace{7pt}
{\hoch{\dagger} \it Department of Physics and Astronomy, Rutgers University,
Piscataway, NJ 08855}

\vspace{7pt}
{\hoch{\star}\umich}

\vspace{7pt}
{\hoch{\ddagger}\tamphys}

\vspace{7pt}
{\hoch{\ddagger}\ihp}

\vspace{14pt}

\underline{ABSTRACT}
\end{center}

    Modifications to the singularity structure of D3-branes that result
from turning on a flux for the R-R and NS-NS 3-forms (fractional
D3-branes) provide important gravity duals of four-dimensional $N=1$
super-Yang-Mills theories.  We construct generalisations of these
modified $p$-brane solutions in a variety of other cases, including
heterotic 5-branes, dyonic strings, M2-branes, D2-branes, D4-branes and
type IIA and type IIB strings, by replacing the flat transverse space
with a Ricci-flat manifold $M_n$ that admits covariantly constant
spinors, and turning on a flux built from a harmonic form in $M_n$, thus
deforming the original solution and introducing fractional branes. The
construction makes essential use of the Chern-Simons or
``transgression'' terms in the Bianchi-identity or equation of motion of
the field strength that supports the original undeformed solution.  If
the harmonic form is $L^2$ normalisable, this can result in a
deformation of the brane solution that is free of singularities, thus
providing viable gravity duals of field theories in diverse dimensions
that have less than maximal supersymmetry. We obtain examples of
non-singular heterotic 5-branes, dyonic strings, M2-branes, type IIA
strings, and D2-branes.

{\vfill\leftline{}\vfill
\footnoterule
{\footnotesize \hoch{1} Research supported in part by DOE grant
DE-FG02-95ER40893 and NATO grant 976951. \vskip -12pt} \vskip 14pt
{\footnotesize \hoch{2} Research supported in full by DOE grant
DE-FG02-95ER40899 \vskip -12pt} \vskip 14pt
{\footnotesize  \hoch{3} Research supported in part by DOE
grant DE-FG03-95ER40917.\vskip  -12pt}}

\pagebreak
\setcounter{page}{1}

\tableofcontents
\vfill\eject

\section{Introduction}

   Chern-Simons type modifications, or ``transgressions,'' in the
Bianchi identities or equations of motion for field strengths are a
commonplace in supergravity theories.  These terms play an important
role in the theory, governing both the bosonic symmetries and the
supersymmetry.  In the usual construction of a $p$-brane solution, one
usually chooses the configuration of non-vanishing field strengths to be
such that the Chern-Simons terms give no contribution in the equations
of motion.\footnote{U-duality transformations can then map the solution
into ones where the Chern-Simons terms play a non-trivial r\^ole, but
these are not the type of solutions that we wish to consider in this
paper, precisely because such solutions are U-duality-related to ones
not involving the Chern-Simons terms.}  In this paper, we obtain new
types of brane solutions where the Chern-Simons type terms play an
essential and intrinsic r\^ole.  These solutions can be viewed as
deformations of the standard brane solutions, in which addition flux is
turned on, and in certain cases the deformation can have the effect of
``resolving'' the singularities of the original solution.  Since the
resolution of such BPS supergravity solutions can break additional
supersymmetry, such non-singular solutions are of special interest,
since they may serve as viable gravity duals of strongly-coupled
Yang-Mills field theories with less than maximal supersymmetry, and thus
may provide important information on these field theories, such as
confinement and chiral-symmetry breaking.

   Our construction applies to the general class of theory where there
is an $n$-form field strength with the Bianchi identity
\be
dF_\n = F_{\sst{(p)}}\wedge F_{\sst{(q)}}\,.
\ee
The field $F_\n$ supports a standard magnetic brane solution\footnote{
In some of our examples, $F_\n$ itself is the Hodge dual of the
``actual'' field strength that appears in the theory, and so it is then
field equation, rather than the Bianchi identity, that receives the
Chern-Simons type correction in such a case.  The field strength in the
theory then carries an electric charge.} with a flat transverse space of
dimension $n+1=p+q$.  One can always replace this transverse space with
any other Ricci-flat space, and provided that it admits
covariantly-constant (sometimes known as ``parallel'') spinors, one can
arrange by suitable orientation choices that the resulting brane
solution will still preserve some supersymmetry.  This replacement of
the transverse-space metric is a key part of the construction that
follows.

    If the transverse space admits a suitable non-trivial harmonic
$p$-form, we can take $F_{\sst{(p)}}$ to be equal to this harmonic form,
with the consequence that the Chern-Simons term can contribute a
non-trivial flux to $F_\n$.  (In some cases $p=q$ and $F_{\sst{(q)}}$ is
the same field $F_{\sst{(p)}}$, while in other cases it can be a
different field.)  In the cases when the harmonic forms are
normalisable, this additional flux can smooth out singularities in the
solution.  If on the other hand the harmonic form is non-normalisable,
then we find that this leads to pathologies in the solutions.  If the
non-normalisablity arises due to a small-radius divergence of the
harmonic function, the deformed brane solution has a (naked)
singularity, whilst if there is a large-radius divergence in the norm,
the solution will not have a well-defined ADM mass.

    Recently, a resolution procedure has been extensively discussed, for
the D3-brane solution of the type IIB theory
\cite{klebtsey,klebstra,ganpol,gubser,zaytse}, since such non-singular
examples may provide important suprgravity dual solutions of
four-dimensional N=1 super-Yang-Mills theory in the
infra-red.\footnote{Related aspects of branes on resolved conifolds were
addressed, e.g., in \cite{gv,kw,vafa}, and the study of non-singular
supergravity gravity duals of N=1 super-Yang-Mills was initiated in
\cite{ps}. See also related work \cite{mn}.} The Bianchi identity for
the self-dual 5-form is given by $dF_\5= -\im\, F_\3\wedge \bar F_\3$,
where $F_\3$ is the complex 3-form whose real and imaginary parts are
the R-R and NS-NS 3-forms of the theory.  The basic idea now is that one
can choose any Ricci-flat six-dimensional manifold $M_6$ for the
transverse space, and if it is additionally K\"ahler, it will admit
covariantly-constant spinors and hence still allow a supersymmetric
D3-brane solution.  Then, one can look for a suitable 3-form in
$M_6$. It turns out that if it is harmonic, and (complex) self-dual, the
type IIB supergravity equations can still be satisfied, providing the
R-R and NS-NS 3-forms equal to the self-dual harmonic 3-form on $M_6$,
thus introducing fractional branes \cite{gp}. Depending upon the
detailed properties of the harmonic form, it may be that the resulting
generalisation of the D3-brane solution becomes regular, thus providing
a valid supergravity solution that is dual to $N=1$ four-dimensional
super-Yang-Mills theory in the infra-red region.  

   In this paper we begin, in section 2, by reviewing the resolved
solutions for the D3-brane.  We also study some aspects of the
specific example of the ``resolved conifold'' introduced in
\cite{zaytse}, which is an alternative deformation of the conifold to
the one discussed previously in
\cite{klebtsey,klebstra,ganpol,gubser}.  In the remainder of the
paper, we generalise the procedure to various other dimensions, and
present a construction of new brane solutions in these cases.

    In section 3 we show how one can replace the 4-dimensional space
transverse to the heterotic 5-brane by any Ricci-flat space $M_4$, and
set a $U(1)$ field chosen from the Yang-Mills sector equal to a
self-dual (or anti-self-dual) harmonic 2-form in $M_4$.  By taking $M_4$
to be K\"ahler, we have the possibility of also partially preserving the
supersymmetry.  We consider several examples, including the cases where
$M_4$ is taken to be Eguchi-Hanson, Taub-NUT or the multi-centre
generalisations of these metrics.  Both the cases of Eguchi-Hanson and
Taub-NUT admit a normalisable self-dual and anti-self-dual harmonic
2-form, respectively, and we find that the resulting solutions give a
non-singular supersymmetric resolutions of the heterotic 5-brane. The
Eguchi-Hanson metric is of further interest because it can be used to
smooth out the sixteen orbifold singularities of $T^4/Z_2$, as an
orbifold construction of the K3 manifold \cite{gibpop,pag}.  Thus our
solution can be viewed as the 5-brane on a K3 manifold, interpolating
between a flat region in the near-orbifold limit of K3 and the curved
region close to an orbifold point.  It is striking that the 5-brane on
the transverse K3 manifold is completely regular and does not require
external source terms, whilst the 5-brane on the transverse Euclidean
space would require a 5-brane source action.

    In section 4 we consider generalised dyonic string solutions in six
dimensions. The construction here is somewhat similar to the heterotic
5-brane, in that the transverse space is again 4-dimensional.  We again
replace the transverse space by a Ricci-flat K\"ahler manifold $M_4$,
and show that again we can obtain generalised solutions if there is a
self-dual or anti-self-dual harmonic 2-form on $M_4$.  Since the
Eguchi-Hanson (Taub-NUT) metric has a non-trivial normalisable self-dual
(anti-self-dual) harmonic 2-form we can obtain explicit non-singular
resolutions of the dyonic string. Interestingly, this mechanism can also
provide a resolution of repulson-type singularities for the case of
tension-less dyonic strings.

    In section 5, we consider generalisations of the M2-brane solution
of M-theory, in which the 8-dimensional transverse space is replaced by
a Ricci-flat 8-manifold $M_8$.  There are several types of such
manifolds $M_8$ that admit covariantly constant spinors, namely
hyper-K\"ahler, with $Sp(2)$ holonomy; K\"ahler, with $SU(4)$ holonomy;
and then the exceptional case in Berger's classification \cite{berger},
with Spin(7) holonomy.  These admit 4, 2 and 1 covariantly-constant
spinors respectively, and these numbers reflect themselves in the
fractions of preserved supersymmetry in the M2-brane solutions.  We find
that if $M_8$ admits a suitable self-dual (or anti-self-dual) harmonic
4-form, one can add this to the 4-form field strength, and still obtain
a solution of the eleven-dimensional equations of motion.  This gives
generalisations of the M2-brane solution, which would, for a suitable
harmonic 4-form, give rise to a non-singular resolution of the original
M2-brane.  We discuss an example in detail, for an $M_8$ manifold of
Spin(7) holonomy whose Ricci-flat metric was constructed in
\cite{brysal,gibpagpop}.  We find a normalisable harmonic 4-form, and we
then use this to construct a resolved M2-brane solution that is somewhat
analogous to the one we found for the heterotic 5-brane with the
Eguchi-Hanson transverse metric.

   In section 6 we consider generalisations of the D2-brane solution of
the type IIA theory.  These involve replacing the 7-dimensional flat
transverse space by a Ricci-flat 7-manifold $M_7$.  We show that if
$M_7$ admits an harmonic 3-form $L_\3$, then one can construct
generalised D2-brane solutions in which the 3-form $F_\3$ is taken to be
proportional to $L_\3$, while a term proportional to ${*L_\3}$ is added
to the 4-form $F_\4$.  (Here $*$ denotes the Hodge dual in the
7-dimensional metric on $M_7$.)  Cases that admit covariantly-constant
spinors include the exceptional manifolds of $G_2$ holonomy occurring in
the Berger classification \cite{berger}.  We construct an example of a
generalised D2-brane, using a simple 7-manifold of $G_2$ holonomy.  This
particular example yields a linearly non-normalisable harmonic form in
the asymptotic region, and the corresponding D2-brane is non-singular
everywhere, but the function $H$ does not fall off fast enough
asymptotically to give a well-defined ADM mass in this case.

   In section 7, we discuss a few further examples of deformed brane
solutions.  Specifically, we construct string solutions in the type IIA
and type IIB theories, and D4-branes in the type IIA theory.  The type
IIA strings are nothing but diagonal dimensional reductions of the
M2-branes that we construct in section 5, and so they are again
supported by harmonic 4-forms in the 8-dimensional Ricci-flat transverse
manifold.  By contrast, the mechanism for obtaining deformed string
solutions in the type IIB theory is rather different, and in this case
we find that they arise if the 8-manifold has an harmonic 3-form.
Finally, the D4-brane solutions involve a Ricci-flat 5-dimensional
transverse space, with an harmonic 2-form.

   In section 8 we summarise the results, and comment on possible
generalisations and further studies in connection with field-theory
duals.

\section{D3-branes on Ricci-flat K\"ahler 6-manifolds}

\subsection{General discussion}

    We begin with a brief review of the deformed D3-brane solutions,
which have been discussed extensively in recent times 
\cite{klebtsey,klebstra,ganpol,gubser,zaytse}.

      The bosonic sector of ten-dimensional type IIB supergravity
comprises the metric, a self-dual 5-form field strength, a scalar, an
axion, an R-R 3-form and an NS-NS 3-form field strength.  There is no
simple covariant Lagrangian for type IIB supergravity, on account of the
self-duality constraint for the 5-form.  However, one can write a
Lagrangian in which the 5-form is unconstrained, which must then be
accompanied by a self-duality condition which is imposed by hand at the
level of the equations of motion \cite{bergs}.  This type IIB Lagrangian
is
\bea
{\cal L}^{\rm IIB}_{10} &=& \hat R\, {\hat *\oneone} - \ft12
{\hat *d{\phi}}\wedge d{\phi} - \ft12
e^{2{\phi}}\,
{\hat *d{\chi}}\wedge d{\chi} - \ft14 { \hat *F_\5}
\wedge F_\5 \nn\\
& & -\ft12 e^{-\phi} \, {\hat *F^{\ns}_\3}\wedge {F}^{\ns}_\3 -
\ft12 e^{\phi}\, {\hat *{F}^{\rr}_\3}\wedge {F}^{\rr}_\3 -
\ft12 B_\4\wedge d{A}_\2^{\rr}\wedge d{A}_\2^{\ns}\ ,\label{d102blag}
\eea
where $F^\ns_\3 = dA^\ns_\2,\, F^\rr_\3=d A^\rr_\2 - {\chi}\,
d{A}^\ns_\2,\, {F}_\5=d{B}_\4 - \ft12 {A}_\2^\rr\wedge d{A}_\2^\ns +
\ft12 {A}_\2^\ns\wedge d{A}_\2^\rr$.  Note that $A_\2^\rr$ and
$A_\2^\ns$ are the R-R and NS-NS 2-form potentials respectively.  Our
convention here, and throughout the paper, is that the metric in the
theory that we are considering will be denoted by $d\hat s^2$, and
likewise all quantities associated with it, such as the Ricci tensor
$\hat R$, will carry hats.  In particular, $\hat *$ in the present case
denotes the ten-dimensional Hodge dual.

      One can then make the following ``deformed'' D3-brane Ansatz, 
\bea
d\hat s_{10}^2 &=& H^{-1/2}\, dx^\mu\, dx^\nu\, \eta_{\mu\nu} +
H^{1/2}\, ds_6^2\,,\nn\\
F_\5 &=& d^4 x\wedge dH^{-1} + {\hat * dH}\,,\qquad
F_3 \equiv F_\3^{\rr} + {\rm i}\, F_\3^{\ns} = m\, L_\3\,,
\label{d3braneansatz}
\eea
where $ds_6^2$ is any six-dimensional Ricci-flat K\"ahler metric
that admits a non-trivial complex harmonic self-dual 3-form $L_\3
={\rm i}\, {*L_\3}$, and $*$ is the Hodge dual with respect to
$ds_6^2$.  Again, our notation here and throughout the paper is
that the unhatted metric always denotes the one in the space
transverse to the brane world-volume, and $*$ denotes the Hodge
dual with respect to this metric.

     It is straightforward to verify the following:
\bea
d{\hat *F}_\3= {\rm i}\, F_\5\wedge F_\3 &\Rightarrow&
{\rm i}\, {* L_\3} = L_\3\,,\qquad d L_\3=0\,,\nn\\
{\hbox{$(\chi, \phi)$ equations}} &\Rightarrow&
{\rm i}\, {* L_\3} = L_\3\,,\nn\\
dF_\5 = -{\rm i}\,  F_\3\wedge \bar F_\3 &\Rightarrow&
\square H = -\ft1{12} m^2\,  |L_\3|^2\,,\\
{\hbox{Einstein equation}} &\Rightarrow&
\square H = -\ft1{12} m^2\, |L_\3|^2\,,\nn
\eea
where
\be
|L_\3|^2 \equiv L_{mnp}\, \bar L^{mnp}\,,
\ee
and $\square$ denotes the scalar Laplacian calculated in the
six-dimensional transverse-space with metric $ds_6^2$.
Thus all the equations of motion are satisfied provided that
\be
\square H = -\ft1{12} m^2\, |L_\3|^2\,.\label{d6heq}
\ee
Note that there is a correlation between the sign in the duality
equation $\im\, {*L_\3}=L_\3$, and the sign of the Chern-Simons or
transgression term in the type IIB theory.

   We see that the complex self-dual harmonic 3-form $L_\3$ acts as a
source for the function $H$.  In Euclidean 6-space, there is always a
simple harmonic form 3-form for which $L_\3\wedge \bar L_\3$ becomes the
volume form.  This is not especially interesting since it contributes a
term $-m^2\, r^2$ to the function $H$, which implies that the solution
contains a naked singularity.  More interesting solutions can be
obtained by taking the transverse space to be some non-trivial complete
non-compact Ricci-flat 6-manifold.  An example that has been much
studied is the ``deformed conifold'' that was introduced in
\cite{candel}, and studied further in \cite{mintsi,ohyo}.  The
deformed D3-brane solution using this metric  was constructed in
\cite{klebstra}.  In section 2.2, we shall study some aspects of
another deformed D3-brane solution that has recently been discussed,
using a different complete Ricci-flat 6-manifold \cite{zaytse}.

\subsection{D3-brane on a Ricci-flat K\"ahler 6-manifold}

    The transverse 6-metric that we shall consider here is the one
discussed recently in \cite{zaytse}, where it was used to obtain a
deformed D3-brane solution with NS-NS and R-R flux.
It is a metric of cohomogeneity one, whose level
surfaces are the 5-dimensional manifold of the $U(1)$ bundle over
$S^2\times S^2$, where the $U(1)$ fibre has winding number 1 over each
2-sphere.  The metric can be written as
\be
ds_6^2 = h^2\, dr^2 + \a^2\, \sigma^2 + \beta^2\, d\Omega_2^2 +
\gamma^2\, d\wtd\Omega_2^2\,,\label{stenzel}
\ee
where $h$, $\a$, $\beta$ and $\gamma$ are functions only of $r$, and
\bea
&&d\Omega_2^2 = d\theta^2 + \sin^2\theta\, d\phi^2 \,,\qquad
d\wtd\Omega_2^2 = d\td\theta^2 + \sin^2\td\theta\, d\td\phi^2 \,,\nn\\
&&\sigma = d\psi +\cos\theta\, d\phi + \cos\td\theta\,
d\td\phi\,.\label{oms}
\eea
Choosing $e^0=h\, dr$, $e^1 = \beta\, d\theta$,
$e^2= \beta\, \sin\theta\, d\phi$,  $e^3 = \gamma\, d\td\theta$,
$e^4= \gamma\, \sin\td\theta\, d\td\phi$, $e^5=\a\, \sigma$ for the
orthonormal frame, the
tangent-space components of the Ricci tensor are given by
\bea
R_{00} &=& -\fft{1}{h\, \alpha} \, \Big(\fft{\alpha'}{h}\Big)'
              -\fft{2}{h\, \beta} \, \Big(\fft{\beta'}{h}\Big)'
                -\fft{2}{h\, \gamma} \, \Big(\fft{\gamma'}{h}\Big)'
\,,\nn\\
R_{11}&=&R_{22} = -\fft{1}{h\, \beta} \, \Big(\fft{\beta'}{h}\Big)'
         -\fft{{\beta'}^2}{h^2\, \beta^2}
    - \fft{2\beta'\, \gamma'}{h^2\, \beta\, \gamma}
   -\fft{\a'\, \beta'}{h^2\, \a\, \beta} + \fft1{\beta^2}
       -\fft{\a^2}{2\beta^4}\,,\nn\\
R_{33}&=&R_{44} = -\fft{1}{ h\, \gamma} \, \Big(\fft{\gamma'}{h}\Big)'
         -\fft{{\gamma'}^2}{h^2\, \gamma^2}
    - \fft{2\beta'\, \gamma'}{h^2\, \beta\, \gamma}
   -\fft{\a'\, \gamma'}{h^2\, \a\, \gamma} + \fft1{\gamma^2}
       -\fft{\a^2}{2\gamma^4}\,,\nn\\
R_{55} &=& -\fft{1}{ h\, \a} \, \Big(\fft{\a'}{h}\Big)'
           +\fft{\a^2}{2\beta^4} +\fft{\a^2}{2\gamma^4}
    -\fft{2\a'\, \beta'}{h^2\, \a\, \beta}
         -\fft{2\a'\, \gamma'}{h^2\, \a\, \gamma}\,.\label{stenric}
\eea
It is then easily verified that the following gives a Ricci-flat
metric \cite{zaytse}:
\be
h^2 =  \fft{r^2+6a^2}{r^2+9 a^2} \,,\qquad \a^2 =
\ft19 \Big(\fft{r^2+9a^2}{r^2+6 a^2}\Big)\,  r^2\,, \qquad \beta^2 = \ft16
r^2\,,\qquad \gamma^2 = \ft16 (r^2+ 6 a^2)\,,\label{funsol}
\ee
where $a$ is a constant.  The radial coordinate runs from $r=0$ to
$r=\infty$.  Near $r=0$, the metric smoothly approaches flat $R^4$
times a 2-sphere of radius $a$, while at large $r$ the metric
describes the cone with level surfaces that are the $U(1)$ bundle over
$S^2\times S^2$.  If the parameter $a$ is set to zero, the cone metric
becomes singular at $r=0$, and the manifold degenerates to the
conifold described in detail in \cite{candel}.  The construction of
the homogeneous metrics on the $U(1)$ bundle over $S^2\times S^2$ that
is used here was given in \cite{pagpop}.

    Substituting (\ref{funsol}) back into the expressions for the
curvature, one finds that the curvature 2-forms $\Theta_{ab} =\ft12
R_{abcd}\, e^c\wedge e^d$ are given by
\bea
&&\Theta_{01} = A\, (e^0\wedge e^1 + e^2\wedge e^5)\,,\qquad
 \Theta_{02} = A\, (e^0\wedge e^2 - e^1\wedge e^5)\,,\nn\\
&&\Theta_{03} = -B\, (e^0\wedge e^3 + e^4\wedge e^5)\,,\qquad
 \Theta_{04} = -B\, (e^0\wedge e^4 - e^3\wedge e^5)\,,\nn\\
&&\Theta_{05} = \fft{72 a^4}{(r^2+ 6a^2)^3}\,
(e^0\wedge e^5 + e^3\wedge e^4) - \fft{6 a^2}{(r^2+6a^2)^2}\,
(e^1\wedge e^2 - e^3\wedge e^4)\,,\nn\\
&&\Theta_{12} =\fft{r^2}{r^2+6a^2}\, (e^1\wedge e^2 - e^3\wedge e^4) -
\fft{6a^2}{(r^2+6a^2)^2}\, (e^0\wedge e^5 + e^3 \wedge e^4)\,,\nn\\
&&\Theta_{13} = -C\, (e^1\wedge e^3 + e^2\wedge e^4)\,,\qquad
 \Theta_{14} = -C\, (e^1\wedge e^4 - e^2\wedge e^3)\,,\nn\\
&&\Theta_{15} = A\, (e^1\wedge e^5 - e^0\wedge e^2)\,,\qquad
 \Theta_{23} = -C\, (e^2\wedge e^3 - e^1\wedge e^4)\,,\nn\\
&&\Theta_{24} = -C\, (e^2\wedge e^4 + e^1\wedge e^3)\,,\qquad
 \Theta_{25} = A\, (e^2\wedge e^5 + e^0\wedge e^1)\,,\nn\\
&&\Theta_{35} = -B\, (e^3\wedge e^5 - e^0\wedge e^4)\,,\qquad
 \Theta_{45} = -B\, (e^4\wedge e^5 + e^0\wedge e^3)\,,
\eea
where we have defined
\be
A\equiv \fft{3a^2}{(r^2+6a^2)^2}\,,\qquad B\equiv
\fft{3a^2\, (r^2+18a^2)}{(r^2+6a^2)^3}\,,\qquad C\equiv
\fft{r^2+9a^2}{(r^2+6a^2)^2}\,.
\ee
The metric is K\"ahler, with K\"ahler form given by
\be
J= e^1\wedge e^2 + e^3\wedge e^4 -e^0\wedge e^5\,.\label{steka}
\ee

   Since the integrability condition for the existence of a
covariantly-constant spinor is $R_{abcd}\, \Gamma^{cd}\, \eta=0$, we
can immediately deduce from the expressions for the curvature 2-forms
that $\eta$ must satisfy
\be
\Gamma_{12}\,\eta = \Gamma_{34}\, \eta = -\Gamma_{05}\, \eta\,.
\label{etacon}
\ee
It is then straightforward to substitute back into the covariant-constancy
equation itself, $D_a\, \eta=0$, to deduce that $\eta$ is given by
\be
\eta = e^{-\fft12\psi\, \Gamma_{12}}\, \eta_\0\,,\label{ks2}
\ee
where $\eta_\0$ is a constant spinor that satisfies the same
conditions (\ref{etacon}).  There are two such independent solutions.

    As in \cite{zaytse}, one may note that the following is an harmonic
3-form:
\be
\omega_\3 = \sigma\wedge \Omega_\2 - \sigma\wedge \wtd\Omega_\2\,,
\ee
where $\Omega_\2 \equiv \sin\theta\wedge d\phi$ and
$\wtd\Omega_\2 \equiv \sin\td\theta\wedge d\td\phi$ are the volume
forms on the two unit 2-spheres.  The Hodge dual of $\omega_\3$ is
given by
\be
{*\omega_\3} = \fft{h\, \gamma^2}{\a\, \beta^2}\, dr\wedge
\wtd\Omega_\2 - \fft{h\, \beta^2}{\a\, \gamma^2}\, dr\wedge
\Omega_\2\,.
\ee
Thus the closure and co-closure of $\omega_\3$ is manifest.

    It is clear that from $\omega_\3$ we can construct the complex
self-dual harmonic 3-form
\be
L_\3 = \omega_\3 + \im\, {*\omega_\3}\,,\label{l3def}
\ee
satisfying $L_\3 = \im\, {*L_\3}$.  It is then easy to see that
\be
|L_\3|^2 = \fft{7776(18a^4 + 6 a^2\, r^2 + r^4)}{r^6\, (r^2+6 a^2)
(r^2 + 9 a^2)}\,.
\ee
This harmonic 3-form is not normalisable, owing to the strength of its
divergence as $r\rightarrow 0$, and so we can expect that the
deformed D3-brane will have a singularity for small $r$.  Furthermore,
its fall-off at large $r$ is insufficient to give normalisability there;
there will be a logarithmic divergence.

    If we assume the function $H$ depends only on $r$, then
(\ref{d6heq}) becomes
\be
\Big( r^3\, (r^2 + 9 a^2)\, H'\Big)' = -\fft{648 m^2\, (18 a^4 + 6
a^2\, r^2 + r^4)}{r^3\, (r^2 + 9 a^2)}\,.
\ee
The first integration of this equation gives
\be
 r^3\, (r^2 + 9 a^2)\, H' = -81 b + \fft{648 a^2\, m^2}{r^2} -
288 m^2\, \log r - 180 m^2\, \log(r^2+9 a^2)\,,
\ee
and the second gives
\bea
H &=& 1 + \fft{3b - 4m^2 + 24 m^2\, \log (3a)}{3 a^4}\, \log r
+\fft{4m^2 - 3 b}{6 a^4}\, \log(r^2+9a^2)  \nn\\
&&+\fft{m^2}{9 a^4}\, (4 \log r -5 \log(r^2+9a^2))(4 \log r +
\log(r^2+9 a^2))\\
&&+ \fft{9b + 24 m^2 + 32 m^2\, \log r + 20 m^2
\log(r^2+ 9a^2)}{2a^2\, r^2}-\fft{18 m^2}{r^4} -
\fft{2m^2}{a^4}\, \hbox{Li}_2\Big(-\fft{r^2}{9a^2}\Big)\,,\nn
\eea
where Li$_2(x) = \int_x^0 dt \log(1-t)/t$ is the dilogarithm function,
and $b$ is a constant of integration.  (This is the explicit form of the
solution whose general structure was discussed in \cite{zaytse}.)  The
leading-order behaviour for $H$ is given by
\bea
r>>a: && H = 1 + \fft{81(b + 2m^2)}{4r^4} + \fft{162m^2\, \log r}{r^4}
+\cdots   \,,\\
r<<a:&& H = 1  -\fft{18m^2}{r^4} + \fft{9b+24m^2+40 m^2\, \log(3a)
      + 32m^2\, \log r}{2 a^2\, r^2} +\cdots
\,.\nn
\eea
Thus the solution has a repulson-type singularity \cite{zaytse}.
Furthermore the logarithmic behaviour of $H$ at large $r$ implies
that the metric does not have a well-defined ADM mass.
These behaviours are a direct consequences of the non-normalisability of
$L_\3$.  Namely, since  $L_\3$ is not normalisable for small $r$, the
solution is singular near the origin; and since it is not normalisable
at large $r$, $H$ will fall off too slowly to have a 
well-defined ADM mass.

     As mentioned earlier, it is also possible to consider a different
Ricci-flat 6-metric, namely the one discussed in the deformed D3-brane
in \cite{klebstra}.  In this case the 6-metric is quite distinct from
the metric of \cite{zaytse} that we have been studying here.  They are
both complete, but the Ricci-flat 6-metric used in \cite{klebstra} has
a minimal $S^3$ in the centre, whilst the Ricci-flat metric in
\cite{zaytse} has a minimal $S^2$ in the centre.  In the D3-brane
solution using this ``deformed conifold'' that was constructed in
\cite{klebstra}, the harmonic 3-form $L_\3$ is normalisable at small
$r$, but logrithmically non-normalisable for large $r$. As a
consequence, the function $H$ has no singularity at small $r$, but has
the same logrithmic behavior at large $r$ as in the solution of \cite{zaytse}.

    The two covariantly-constant spinors, given by (\ref{etacon}) and
(\ref{ks2}), would give rise to Killing spinors of the D3-brane
solution if we merely replaced the flat transverse 6-space of a
standard D3-brane solution by the Ricci-flat manifold with the metric
under discussion here.  Once one turns on the NS-NS and R-R 3-form
flux, by allowing the parameter $m$ to be non-zero, it is necessary to
check the additional conditions that now arise.  A discussion of
supersymmetry in such deformed D3-brane solutions was given in
\cite{ganpol,gubser}.  In particular, from the results in
\cite{ganpol} it is necessary for supersymmetry that the 3-form $L_\3$
be purely of type $(2,1)$, with no admixture of $(1,2)$, $(0,3)$ or
$(3,0)$ terms.   Using the K\"ahler form (\ref{steka}), we can verify that
the complex 3-form $L_\3$ defined in (\ref{l3def}) indeed has no purely
holomorphic or antiholomorphic parts, of type $(0,3)$ or $(3,0)$.
However, it does have terms of type $(1,2)$ as well as $(2,1)$, and so
based on the results in \cite{ganpol}, it would seem that the solution
will not be supersymmetric.

\section{Heterotic 5-branes on Ricci-flat K\"ahler 4-manifolds}

\subsection{General discussion}

         The bosonic sector of the ten-dimensional heterotic
supergravity consists of the metric, a dilaton, a 2-form potential
$A_\2$ and the Yang-Mills fields of $E_8\times E_8$ or $SO(32)$.  The
Lagrangian is given by
\be
{\cal L}_{\rm het} = \hat R\, {\hat *\oneone} -
\ft12 {\hat *d\phi}\wedge d\phi -
\ft12 e^{-\phi}\,  {\hat *F_3}\wedge F_\3
-\ft12 e^{-\ft12\phi}\, {\hat * F_\2^i}\wedge F_\2^i\,,
\ee
where
\bea
F_\3 &=& dA_\2 + \ft12 A_\1^i \wedge dA_\1^i +\ft16
f_{ijk}\, A_\1^i\wedge A_\1^j\wedge A_\1^k\,,\nn\\
F_\2^i &=& dA_\1^i + \ft12 f^i{}_{jk}\,
A_\1^{j}\wedge A_\1^k\,.\label{fielddef}
\eea
The Bianchi identity for $F_\3$ is given by
\be
dF_\3 = \ft12 F_\2^i\wedge F_\2^i\,.\label{hetbianchi}
\ee
This implies that the magnetic 5-brane charge can be supplied by a
Yang-Mills instanton in the 4-space transverse to the heterotic
5-brane.  Such a solution was constructed in \cite{strom}, where it
was shown that the singularity of the standard 5-brane is smoothed out
by the instanton configuration.

   Here, we show that the heterotic 5-brane admits a quite different
kind of deformation, supported by an Abelian $U(1)$ field, again giving
a regular solution, provided that the transverse 4-space admits a
non-trivial self-dual harmonic 2-form.  To do this, we make the
following 5-brane Ansatz,
\bea
d\hat s_{10}^2 &=& H^{-1/4}\, dx^\mu\, dx^\mu\, \eta_{\mu\nu} + H^{3/4}
ds_4^2\,, \nn\\
e^{-\phi}\, {\hat *F_\3} &=& d^6x\wedge dH^{-1}\,,\qquad \phi =
\ft12\log H\,,\qquad F_\2=m\, L_\2\,,\label{het5braneansatz}
\eea
where $L_\2$ is an harmonic 2-form in the Ricci-flat transverse
metric $ds_4^2$.  Note that here we only turn on one of the
Yang-Mills gauge fields, which we write as $F_\2$.  It is
straightforward to verify that the above Ansatz satisfies all the
equations of motion, provided that $L_\2$ is a self-dual harmonic
2-form, (${*L_\2}=L_\2$, $dL_\2=0$) and that
\be
\square H = -\ft14 m^2\, L_\2^2\,.\label{het5branesource}
\ee

    Note that the sign in the duality relation ${*L_\2}=+L_\2$ is
correlated with the sign of the Chern-Simons type terms in the
expression for $F_\3$ in (\ref{fielddef}), and hence in
(\ref{hetbianchi}) too.  It also depends, of course, on our orientation
conventions when taking the Hodge dual.  Consequently, we can perfectly
well also obtain a solution of the above type in a case where we instead
have an anti-self-dual harmonic 2-form in the transverse metric
$ds_4^2$, by making the appropriate orientation change.  One must be
careful, however, when checking the supersymmetry of the solution, since
reversing the orientation of the transverse space can make the
difference between whether or not covariantly-constant spinors exist
that have the required chirality for obtaining supersymmetry in the
generalised 5-brane solution.

   Since $\phi$ and $F_\3$ in the new solutions (\ref{het5braneansatz})
have the same functional dependence on $H$ as they do in the standard
heterotic 5-brane solution, it follows that the gravitino and dilatino
transformation rules will imply very similar conditions for preserved
supersymmetry to those in the standard solution where $ds_4^2$ is flat.
Thus we have
\bea
\delta \psi_M &=& \hat D_M\, \ep - \ft1{96}\, e^{-\fft12\phi}\, F_{NPQ}\,
(\Gamma_M{}^{NPQ} - \delta_M^N\, \Gamma^{PQ})\, \ep =0\,,\nn\\
\delta \lambda &=& \del_M\, \phi\, \Gamma^M\, \ep - \ft1{12}\,
e^{-\fft12\phi}\, F_{MNP}\, \Gamma^{MNP}\, \ep=0\,,
\eea
which implies first of all the usual condition
\be
\ft1{4!}\, \ep_{abcd}\, \Gamma^{abcd}\, \ep + \ep =0\,,
\ee
where the indices $a,b,\ldots$ range over the 4-dimensional
transverse space.  In addition, the gravitino transformation rule
now requires that after decomposing $\ep$ as the product of a
spinor in the six-dimensional brane world-volume and a spinor
$\eta$ in the four-dimensional transverse space, $\eta$ must be
covariantly constant in the metric $ds_4^2$.  This implies that
we must take the Ricci-flat metric $ds_4^2$ to be K\"ahler, and
that the 4-manifold must be oriented appropriately.  Finally, the
gaugino transformation rules
\be
\delta \chi^i = F^i_{MN}\, \Gamma^{MN}\, \ep
\ee
imply that $\eta$ must also satisfy
\be
L_{ab}\, \Gamma^{ab}\, \eta =0\,,\label{gauginocon}
\ee
in order to have supersymmetry.

\subsection{Heterotic 5-brane on Eguchi-Hanson instanton}

      Let us consider the case where the Ricci-flat transverse
4-metric $ds_4^2$ is the Eguchi-Hanson solution \cite{egha},
\bea
ds_4^2 &=& W^{-1}\, dr^2 + \ft14 r^2\, W\, (d\psi+\cos\theta\, d\phi)^2
+ \ft14 r^2\, d\Omega_2^2\,,\nn\\
W &=& 1 - \fft{a^4}{r^4}\,,\label{eguchihanson}
\eea
where $d\Omega_2^2=d\theta^2 + \sin^2\theta\, d\phi^2$.  The radial
coordinate $r$ lies in the range $a\le r \le\infty$, and $\psi$ has
period $2\pi$.  The metric is asymptotically locally Euclidean (ALE),
with the periodicity condition on $\psi$ implying that that the level
surfaces at constant $r$ are $RP^3=S^3/Z_2$.  It is K\"ahler, with
self-dual curvature.  The K\"ahler form, which is anti-self-dual in
these conventions, is given by
\be
J= \ft12 r\, dr\wedge (d\psi+\cos\theta\, d\phi) - \ft14 r^2\,
\Omega_\2 = e^0\wedge e^3 - e^1\wedge e^2\,,\label{ehka}
\ee
where $\Omega\equiv \sin\theta\, d\theta\wedge d\phi$ is the
volume-form of the unit 2-sphere metric $d\Omega_2^2$, and we define
the orthonormal basis
\be
e^0= W^{-1/2}\, dr\,,\quad e^1 = \ft12 r\, d\theta\,,\quad e^2=\ft12 r\,
\sin\theta\, d\phi\,,\quad e^3 = \ft12 r\, W^{1/2}\,
(d\psi+\cos\theta\, d\phi)\,.
\ee
The curvature 2-forms are given by
\bea
&&\Theta_{01} = \Theta_{23} = -\fft{2 a^4}{r^6}\, (e^0\wedge e^1 +
e^2\wedge e^3)\,,\nn\\
&&\Theta_{02} = \Theta_{31} = -\fft{2 a^4}{r^6}\, (e^0\wedge e^2 +
e^3\wedge e^1)\,,\label{ehcurv}\\
&&\Theta_{03} = \Theta_{12} = \fft{4 a^4}{r^6}\, (e^0\wedge e^3 +
e^1\wedge e^2)\,.\nn
\eea
From the integrability condition $R_{abcd}\, \Gamma^{cd}\, \eta=0$ for
covariantly-constant spinors it follows that there are two,
which satisfy the projection condition
\be
(\Gamma_{03} + \Gamma_{12})\, \eta = 0\,.\label{ehspinor}
\ee

   The metric also admits a self-dual harmonic 2-form, given by
\be
L_\2 = r^{-3}\, dr\wedge (d\psi + \cos\theta\, d\phi) + \ft12 r^{-2}\,
\Omega_2 = \fft2{r^4}\, (e^0\wedge e^3 \wedge e^1\wedge e^2)\,.
\label{l2exp}
\ee
The square of $L_\2$ is given by
\be
L_\2^2 = \fft{16}{r^8}\,,
\ee
and so this harmonic 2-form is normalisable.

     Making the assumption that $H$ depends only on $r$, we now find
that (\ref{het5branesource}) becomes
\be
(r^3\, W\, H') = -\fft{4m^2}{r^5}\,.
\ee
The solution for $H$ is given by
\be
H = 1 + \fft{m^2 + a^4\, b}{4a^6} \log(\fft{r^2-a^2}{r^2+a^2})
+\fft{m^2}{2 a^4\, r^2}\,.\label{hxxx}
\ee
where $b$ in an arbitrary integration constant.  (We have chosen the
second (additive) constant of integration to be 1 for convenience.)
Since the coordinate $r$ runs from $a$ to infinity, it follows that in
general there is a naked singularity when $r$ is close to $a$.  However,
we can choose the constant $b=-m^2/a^4$ such that the logarithmic term
cancels, giving
\be
H=1 + \fft{m^2}{2a^4\, r^2}\,.\label{h2}
\ee
Note that in the region where $r>> a$, the Eguchi-Hanson metric is
asymptotically locally Euclidean, and $H$ in (\ref{hxxx}) or
(\ref{h2}) has the usual $1/r^2$ power-law fall-off.  When
$b=-m^2/a^4$,  the 5-brane solution is completely non-singular.

   The supersymmetry of the solution is easily determined.  From the
general discussion in section 3.1, and the condition (\ref{ehspinor})
for covariantly-constant spinors in the Eguchi-Hanson metric, we see
that the gravitino and dilatino transformation rules imply that the
solution will preserve one half of the original supersymmetry.  (Note
that (\ref{ehspinor}) just amounts to a chirality condition on $\eta$,
which is the same as the condition for the usual heterotic D5-brane.)
Furthermore since the harmonic 2-form $L_\2$ in our new solution is
self-dual, given by (\ref{l2exp}), it immediately follows from
(\ref{ehspinor}) that the condition (\ref{gauginocon}) following from
the gaugino transformation rule is satisfied.  Thus the new solution
preserves one half of the original supersymmetry.

   Another comment is in order.  After a dimensional reduction (along
$\{\theta,\phi,\psi\}$ direction), the above smooth solution can be
interpreted as a BPS domain wall solution in D=7 gauged supergravity,
preserving 1/4 of the original supersymmetry.  It is therefore expected
that the equations of motion for this configuration can be written as a
coupled system of first-order differential equations for the scalar
fields and the conformal factor of the conformally flat space-time
metric, which are governed by the specific form of the superpotential of
the scalar fields. (These equations were first discussed in the context
of four-dimensional BPS supergravity domain walls in \cite{cgr}; for a
review see \cite{cs}.)  Note that the derivation of the explicit form of
the superpotential for the case in consideration would yield information
on the dual six-dimensional field theory, but we relegate this
derivation (as well as those of all the subsequent domain-wall examples
in this paper) to further study \cite{clp}.

       Having obtained the supersymmetric smooth heterotic 5-brane on
the Eguchi-Hanson metric, it is of interest to study the spectrum of a
minimally-coupled scalar in this gravitational background, since this
provides an information on the (``glue-ball'') spectrum in the infra-red
regime of the dual N=2 six-dimensional field theory (see, for example,
\cite{cglp} and references therein). The equation is of the form
\be
\fft{1}{\sqrt{\hat g}}\del_{\sst M}\,
(\sqrt{\hat g}\, \hat g^{\sst{MN}}\, \del_{\sst N} \chi)=0\,.
\ee
In a suitable decoupling limit, the 1 in the function $H$ in
(\ref{h2}) can be dropped. Making an Ansatz for $\chi$ with $\chi
=e^{{\rm i} p\cdot x}\, H^{1/2} W^{-1/4}\, \psi(r)$, and
performing a coordinate transformation $r^4= a^4/(\tanh(2z) -1)$,
where $z$ runs from 0 to infinity, the wave equation becomes
\be
(-\del_z^2-V)\, \psi= \fft{m^2\, p^2}{2a^2}\, \psi\,,
\ee
where the Schr\"odinger potential $V$ is given by
\be
V = \fft{\cosh(4z) -3}{2\sinh^2(2z)}\,.
\ee
The potential approaches $-1/(4z^2)$ as $z\rightarrow 0$, whilst
it approaches a positive constant, 1, for large $z$.    In spite
of the attractive potential as $z\to 0$, it turns out  that the
boundary conditions eliminate the non-positive energy
bound-states  and the spectrum turns out to be  continuous, with
a mass gap $2a^2/m^2$.

      The Eguchi-Hanson metric is particular interesting because it can
be used to smooth out the sixteen orbifold singularities of $T^4/Z_2$,
as an orbifold construction of the K3 manifold \cite{gibpop,pag}.  Thus
our solution can be viewed as the 5-brane on K3 manifold, interpolating
between a flat region in the near-orbifold limit of K3 and the curved
region close to an orbifold point.  It is striking that the 5-brane on
the transverse K3 manifold is completely regular and does not require
external source term, whilst the 5-brane on the transverse Euclidean
space would require 5-brane action source.  It is worth noting that the
number of orbifold singularities of $T^4/Z_2$ is precisely the same as
the number of Cartan generators of the heterotic string theories.

   Note that if we take the limit where the Eguchi-Hanson scale-size $a$
tends to zero, so that $ds_4^2$ becomes locally Euclidean 4-space, the
self-dual $L_\2$ will not contribute any flux.  Making convenient
choices for the integration constants, $H$ is then given by
\be
H = 1 + \fft{Q}{r^2} - \fft{m^2}{3 r^6}\,.
\ee
Thus in this flat-space limiting case of the Eguchi-Hanson metric,
the inclusion of the $L_\2$ term now leads to a naked singularity that
can no longer be removed.

   We could instead have chosen the harmonic 2-form $L_\2$ to be the
K\"ahler form (\ref{ehka}), which is anti-self-dual.  This is not
normalisable, and in fact if we take $L_\2=J$ we shall have $L_\2^2=4$.
As we discussed previously, we can still use this anti-self-dual
harmonic 2-form to construct a generalised 5-brane solution, provided
that we first reverse the orientation of the Eguchi-Hanson manifold so
that it becomes self-dual.  Substituting into (\ref{het5branesource}),
we then find that $H$ is given by
\be
H =  1 -\ft18 m^2\, r^2 + \fft{(a^4\, m^2 - 4b)}{16 a^2}\,
\log\Big(\fft{r^2+a^2}{r^2-a^2}\Big)\,,
\ee
implying an unavoidable singularity.  Furthermore, it is now evident
from the criterion (\ref{gauginocon}) for supersymmetry, which comes
from the gaugino transformation rule, that the anti-self-duality of
$L_\2$ will conflict with (\ref{ehspinor}), and so in this solution
there would be no supersymmetry.

\subsection{Heterotic 5-brane on Taub-NUT instanton}

    Another example of a Ricci-flat K\"ahler 4-metric is the
Taub-NUT instanton \cite{hawk},
\be
ds_4^2 = \Big(\fft{r+a}{r-a}\Big)\, dr^2 + 4a^2\, \Big(\fft{r-a}{r+a}\Big)
(d\psi + \cos\theta\, d\phi)^2 + (r^2-a^2)\, (d\theta^2 +
\sin^2\theta\, d\phi^2)\,,
\ee
where the radial coordinate runs from $r=a$ to $r=\infty$, and $\psi$
has period $4\pi$.  Topologically, the Taub-NUT manifold is $\R^4$, but
although the metric at large $r$ is asymptotically flat, it approaches
the cylinder $\R^3\times S^1$ rather than Euclidean space.  In the
obvious orthonormal frame
\bea
&&e^0= \Big(\fft{r-a}{r+a}\Big)^{-1/2}\, dr\,,\quad e^1=
(r^2-a^2)^{1/2}\, d\theta\,, \qquad e^2 = (r^2-a^2)^{1/2}\,
\sin\theta\, d\phi\,,\nn\\
&&e^3 = 2a\, \Big(\fft{r-a}{r+a}\Big)^{1/2}\,
(d\psi + \cos\theta\, d\phi)\,,
\eea
the metric is anti-self-dual, with the curvature 2-forms given by
\bea
&&\Theta_{01} = -\Theta_{23} = \fft{a}{(r+a)^3}\, (-e^0\wedge e^1 +
e^2\wedge e^3)\,,\nn\\
&&\Theta_{02} = -\Theta_{31} = \fft{a}{(r+a)^3}\, (-e^0\wedge e^2 +
e^3\wedge e^1)\,,\label{tncurv}\\
&&\Theta_{03} = -\Theta_{12} = \fft{2a}{(r+a)^3}\, (e^0\wedge e^3 -
e^1\wedge e^2)\,.\nn
\eea
The integrability condition for covariantly-constant spinors therefore
implies
\be
(\Gamma_{03}-\Gamma_{12})\, \eta = 0\,.\label{taubeta}
\ee

   We now make the following Ansatz for a potential $B_\1$ for a
harmonic 2-form $G_\2=dB_\1$:
\be
B_\1 = f\, (d\psi + \cos\theta\, d\phi)\,,\label{b1ans}
\ee
where $f$ is a function only of $r$.  This gives the field strength
\be
G_\2= f'\, dr\wedge (d\psi+\cos\theta\, d\phi) - f\,
\sin\theta\, d\theta\wedge d\phi = \fft{f'}{2a}\, e^0\wedge e^3 -
\fft{f}{r^2-a^2}\, e^1\wedge e^2\,.
\ee
Imposing self-duality or anti-self-duality (and thus ensuring that
$G_\2$ will be harmonic), we find
\be
f=f_+\equiv  \fft{r+a}{r-a}\,,\qquad \hbox{or}\qquad f=f_-\equiv
\fft{r-a}{r+a}\,,\label{fpm}
\ee
respectively.

    The anti-self-dual choice gives a regular 2-form $G_\2^-$, for which
\be
(G^-_\2)^2 = \fft{4}{(r+a)^4}\,.
\ee
Clearly $G^-_\2$ is normalisable, and so in view of the fact that the
Taub-NUT manifold is topologically $\R^4$, and so trivial, it follows
that $G^-_\2$ must in fact be an exact 2-form.  Indeed, we see from
(\ref{b1ans}) that if $f=(r-a)/(r+a)$ then $B_\1$ is globally defined,
since the coefficient of $(d\psi+\cos\theta\, d\phi)$ tends
appropriately to zero as $r$ approaches $a$.  (Note, however, that
$B_\1$ itself is not normalisable.)

    Solving for the function $H$ using $L_\2=G^-_\2$
(\ref{het5branesource}) (after first reversing the orientation of the
Taub-NUT manifold so that this normalisable harmonic 2-form becomes
self-dual and thus satisfies the heterotic equations of motion in the
conventions that we are using), we obtain
\be
H = 1  - \fft{4 a\, b\, + m^2}{4 a\, (r-a)} + \fft{m^2}{4a\, (r+a)}\,.
\ee
If we choose the integration constant $b$ so that $b=-m^2/(4a)$,
then the function $H$ becomes non-singular in the entire radial
coordinate range, $a\le r\le \infty$.  The $1/r$ behaviour of the
function $H$ at large $r$ is explained by the fact that the
Taub-NUT metric approaches the cylinder $\R^3\times S^1$.  Since
$G^-_\2$ is anti-self-dual, we see that (\ref{gauginocon}) is
compatible with the condition (\ref{taubeta}) on the
covariantly-constant spinors, and so the solution will preserve
half the original supersymmetry.\footnote{Lest there be confusion
about orientation conventions here, we should emphasise again
that in order to fit in with the conventions we adopted for the
heterotic theory and the Ansatz (\ref{het5braneansatz}) in this
paper, which requires that $L_\2$ be self-dual for the solution,
we would need to reverse the orientation of the Taub-NUT metric
relative to the one given above in which its curvature was
anti-self-dual (\ref{tncurv}). This would change the condition
(\ref{taubeta}) on the covariantly-constant spinor to
$(\Gamma_{03}+\Gamma_{12})\, \eta=0$, and the crucial point is
that this is compatible with the gaugino condition
(\ref{gauginocon}) for supersymmetry, $L_{ab}\, \Gamma^{ab}\,
\ep=0$. In order to try to avoid a tedious and repeated
re-discussion of this basic issue in later parts of the paper, we
shall sometimes tend to speak of using an harmonic form of the
``wrong'' duality in a solution without labouring the point that
one would first need to reverse the orientation of the transverse
metric.}

   The other possibility is to take $f=f_+$ in (\ref{b1ans}), in which
case we get the self-dual harmonic 2-form
\be
G_\2^+ = -\fft1{(r-a)^2}\, (e^0\wedge e^3 + e^1\wedge e^2)\,.
\ee
This has $(G_\2^+)^2= 4/(r-a)^4$ and so it is clearly
non-normalisable.  Substituting $L_\2=G_\2^+$ into
(\ref{het5branesource}) we now obtain
\be
H = 1 - \fft{a\, m^2}{3(r-a)^3} - \fft{m^2}{2(r-a)^2} -
\fft{b}{r-a}\,,
\ee
and so the solution is clearly singular.  Furthermore, since $L_\2$ is
now self-dual, it follows that the gaugino criterion (\ref{gauginocon})
for supersymmetry is incompatible with (\ref{taubeta}), and so this
solution does not preserve any supersymmetry.

\subsection{Heterotic 5-branes on multi Eguchi-Hanson and Taub-NUT
instantons}

   We may also consider the multi Eguchi-Hanson or Taub-NUT metrics
\cite{gibhaw},
\be
ds_4^2 = V^{-1}\, (d\tau + A_i\, dx_i)^2 + V\, dx_i\, dx_i\,,
\ee
where $\del_i\, \del_i\, V=0$ and the curvature is self-dual if we take
\be
\del_i \,V = \ep_{ijk}\, \del_j\, A_k\,.
\ee
This solution can describe both $N$ Eguchi-Hanson or $N$ Taub-NUT
instantons, where $V$ is chosen as follows:
\bea
\hbox{$N$ Eguchi-Hanson}:&& V = \sum_{\a=1}^{N+1} \fft{1}{|\vec x-\vec
x_\a|}\,,\nn\\
\hbox{$N$ Taub-NUT}: && V = 1 + \sum_{\a=1}^N
\fft{1}{|\vec x-\vec x_\a|}
\,.
\eea

      Let us choose the orthonormal frame $e^0 = V^{-1/2}\, (d\tau +
A_i\, dx_i)$, $e^i = V^{1/2}\, dx_i$, and make the Ansatz
\be
B_\1 = f\, (d\tau + A_i\, dx_i)
\ee
for the potential $B_\1$ for an harmonic 2-form $G_\2= dB_\1$, where
$f$ depends only on the three $x_i$ coordinates.  Then we find
\bea
G_\2 &=& \del_i \, f\, dx_i\wedge (d\tau + A_i\, dx_i) + \ft12 f\,
\ep_{ijk}\, \del_k\, V\, dx_i\wedge dx_j\,,\nn\\
&=& \del_i\, f\, e^0\wedge e^i + \ft12 f\, V^{-1}\, \del_k\, V\,
\ep_{ijk}\, e^i\wedge e^j\,.
\eea
From this, it follows that $G_\2$ will be self-dual or anti-self-dual
(and hence it will be harmonic, since we already know that $dG_\2=0$) if
\be
f= V \,,\qquad \hbox{or}\qquad f= V^{-1}\,,
\ee
respectively.  Since we have chosen conventions so that the curvature
to be self-dual, it follows that the K\"ahler form will be
anti-self-dual.   For the two cases, the equation for $H$ in
(\ref{het5branesource}) becomes
\bea
f=V:&& \del_i\del_i H = - m^2\, V\, (\del_i V)^2\,.\nn\\
f=\fft{1}{V}:&& \del_i\del_i H = -\fft{m^2}{V^3}(\del_i V)^2\,.
\eea
(We refer to footnote 4 for the
explanations associated with using the anti-self-dual 2-form.) The
solutions are given by
\bea
f=V:&& H = c_0 + c_1\, V - \ft1{6}m^2\, V^3\,,\nn\\
f=\fft{1}{V}:&& H = c_0 + c_1\, V - \fft{m^2}{2V}\,,
\eea
where $c_0$ and $c_1$ are integration constants.

    Of course the harmonic 2-forms that we have constructed here are by
no means the only ones that can be found.  We can expect that if there
are $N+1$ centres in the harmonic function $V$, then there should be in
total $N$ independent normalisable (localised) self-dual harmonic
2-forms. (In the context of multi-brane solutions wrapping different
supersymmetric two-cycles see \cite{ch}.) In the case of orbifold
construction of K3 manifold, there are sixteen Eguchi-Hanson instantons
and hence sixteen localised self-dual harmonic 2-forms.  It follows that
the most general solution of heterotic 5-branes on K3 can be
constructed, in the $T^4/Z_2$ orbifold limit, as ones located around the
sixteen orbifold fixed point with each of the sixteen Cartan 2-form
field strengths of Yang-Mills fields equal to the localised self-dual
harmonic 2-form of each Eguchi-Hanson instanton, i.e. employing the
Ansatz: $F_{(2)}^i=m_iL_{(2)}^i$.  Further study of explicit solutions
of this type (with both, multi Eguchi-Hanson and Taub-NUT metic) is
under way \cite{clp}.

\section{Dyonic strings on Ricci-flat K\"ahler 4-manifolds}

    The heterotic string admits a compactification to $D=6$ in which the
internal four-dimensional manifold is taken to be K3.  Various different
six-dimensional theories can be obtained, with different Yang-Mills
gauge groups, depending upon precisely how the $SU(2)$-valued spin
connection of the Ricci-flat K\"ahler K3 is embedded in the $E_8\times
E_8$ or $SO(32)$ gauge group of the ten-dimensional theory \cite{sag1}.
There will also be quantum corrections to the six-dimensional effective
action, whose 1-loop structures can be determined by general arguments
based on the necessary anomaly-freedom of the theory.  The resulting
six-dimensional theories are described by $N=1$ supergravity, coupled to
an $N=1$ hypermultiplet and a Yang-Mills multiplet.  The bosonic sector
comprises the metric $\hat g_{\mu\nu}$, a dilaton $\phi$, a 3-form field
strength $F_\3$, and the Yang-Mills fields $G^a_\2$.  The self-dual part
of the 3-form field belongs to the gravity multiplet, while the
anti-self-dual part and the dilaton belong to the hypermultiplet.  The
field equations \cite{sag1}, including the 1-loop terms, in the language
of differential forms that we are using here take the form \cite{llop}
\bea
\hat R_{\mu\nu} &=& \ft12 \del_\mu\phi\, \del_\nu\phi + \ft14
e^{-2\a\phi}\, [F^2_{\mu\nu} - \ft16 F_\3^2\, \hat g_{\mu\nu}]\nn\\
&& + \ft12 (v\,
e^{-\a\phi} + \td v\, e^{\a\phi})\, [ (G^a)^2_{\mu\nu} - \ft18
(G^a_\2)^2\, \hat g_{\mu\nu}]\ ,\nn\\
d{\hat *d\phi} &=& \a\, e^{-2\a\phi}\,
   {\hat *F_\3}\wedge F_\3 + \ft12\a\, (v\,
e^{-\a\phi} - \td v\, e^{\a\phi})\, {\hat *G_\2^a}\wedge G_\2^a\ ,\nn\\
d(e^{-2\a\phi}\,{\hat *F_\3})
   &=& \ft12 \td v\, G_\2^a\wedge G_\2^a\ ,\label{sag}\\
D[(v\,e^{-\a\phi} + \td v\, e^{\a\phi})\, {\hat *G_\2^a}] &=& v\,
e^{-2\a\phi}\, {\hat *F_\3}\wedge G_\2^a + \td v\, F_\3\wedge G_\2^a\ ,\nn
\eea
where $\a=1/\sqrt2$, and $D$ denotes the Yang-Mills-covariant
exterior derivative.  The constants $v$ and $\td v$ are rational
numbers characteristic of the embedding of the $SU(2)$ holonomy
group of K3 manifold in the original $E_8\times E_8$ or $SO(32)$
Yang-Mills gauge group in $D=10$.  The terms associated with $\td
v$ come from 1-loop corrections. The field strength $F_\3$
satisfies the Bianchi identity
\be
dF_\3 =v\, G_\2^a\wedge G_\2^a\,.
\ee

      The theory admits a dyonic string solution with the standard type
of singular harmonic functions in a flat transverse 4-space.
Alternatively, the electric and magnetic string charges can be supplied
by a Yang-Mills instanton living in the transverse 4-space
\cite{dulupo}, thus giving a non-singular solution. Here, we shall show
that there is another way to obtain a non-singular dyonic string
solution, by instead considering a 4-space with a non-trivial self-dual
2-form, and with the charges now supplied by a $U(1)$ Abelian gauge
field contained within the Yang-Mills fields.  We consider the following
Ansatz for the dyonic string
\bea
d\hat s_6^2&=& (H_1\, H_2)^{-1/2}\, (-dt^2+dx^2) +
(H_1\, H_2)^{1/2}\, ds_4^2\,,\nn\\
F_\3 &=& dt\wedge dx\wedge dH_1^{-1} + {*dH_2}\,,\label{dyonansatz}\\
\phi&=&\a \, \log(H_2/H_1)\,,\qquad
G_\2 = m \,L_\2\,,\nn
\eea
Note that here $G_\2$ is an Abelian gauge field taken from the
original Yang-Mills fields $G_\2^\a$.  It is straightforward to
verify that all the equations of motion are satisfied provided
that $L_\2$ is an harmonic self-dual 2-form in the transverse
4-metric $ds_4^2$, and that $H_1$ and $H_2$ satisfy
\be
\square H_1 = -\ft14\td v\, m^2\, L_\2^2 \,,\qquad
\square H_2 = -\ft14 v\, m^2\, L_\2^2\,.
\ee

       The solutions for $H_1$ and $H_2$ are the same form as the ones
we found for the function $H$ for the heterotic 5-brane.  Here, we shall
consider the dyonic string on the Eguchi-Hanson metric
(\ref{eguchihanson}).  The non-singular solutions for $H_1$ and $H_2$
are then given by
\be
H_1 = 1 + \fft{\td v\, m^2}{2a^2\, r^2}\,,\qquad
H_2 = 1+ \fft{v\, m^2}{2a^2\, r^2}\,.
\ee
In the case of the heterotic string compactified on K3 manifold, the
value of $v$ or $\td v$ can be negative such that the dyonic string
becomes massless \cite{dulupo}, which indicates a phase transition
\cite{seiwit,dulupo}.  Such a solution is usually associated with a
naked singularity in the region where $H_1$ or $H_2$
vanishes. Singularities of this type are of the repulson type
\cite{repulson}, and their resolution via an ``\enhancon'' mechanism
\cite{jopepo} was proposed.  Interestingly, our new resolved dyonic
string solution on the Eguchi-Hanson metric avoids the repulson
singularity if $a$ is taken to be sufficiently large, thus providing an
alternative to the resolution via the \enhancon mechanism.  This way of
avoiding the repulson singularity is somewhat similar to the way in
which it can be avoided for the dyonic string supported by a Yang-Mills
instanton of sufficiently large scale size \cite{dulupo,llop}.

     In an appropriate decoupling limit, the constant 1 in the
functions $H_1$ and $H_2$ can be dropped, and the resulting dyonic
string can be dimensionally reduced to give a $D=3$ domain-wall
solution.  For simplicity, let us consider the case where
$H_1=H_2\longrightarrow R^2/r^2$.  It follows that the $D=3$ domain
wall is given by
\be
ds_3^2 = \fft{r^2}{R^2}\, W\, (-dt^2 + dx^2) + \fft{R^2\, dr^2}{r^2}\,.
\ee
Note that the metric is asymptotically AdS$_3$ at $r\rightarrow \infty$,
implying that the solution is supported by a non-trivial scalar
potential in $D=3$ with a fixed point. The metric can be transformed to
the conformally-flat frame $ds_3^2 =e^{2A(z)}(-dt^2 + dx^2 + dz^2)$, by
means of the coordinate transformation $z = (R^2/r)\,
{}_2\!F_1[1/4,1/2;5/4;a^4/r^4]$, and hence $z$ runs from some negative
value $z^*$ to $z=0$.  The Schr\"odinger potential for the
minimally-coupled scalar has the following behaviour
\bea
z\rightarrow z^*:&& V=-\fft{3}{16(z-z^*)^2}\,,\nn\\
z\rightarrow 0:&& V=\fft{3}{4z^2}\,,
\eea
and so the spectrum (describing the dual three-dimensional field
theory in the infra-red) is discrete, with positive-definite
energy. Note that the energy level  separation is governed by the
the ratio $m^2/a^2$.

\section{M2-branes on 8-manifolds}

\subsection{General discussion}

         The bosonic section of eleven-dimensional supergravity
comprises the metric and a 3-form potential, with the Lagrangian given
by
\be
{\cal L}_{\rm M} = \hat R\, {\hat *\oneone} -\ft12
{\hat *F_\4}\wedge F_\4 + \ft16 F_\4\wedge F_\4\wedge A_\3\,,
\ee
where $F_\4=dA_\3$.  The equation of motion for the $A_\3$ is given by
\be
d{\hat *F_\4} = \ft12 F_\4\wedge F_\4\,\label{eomf4}
\ee
The theory admits an M2-brane solution, which has an 8-dimensional
transverse space.  The equation of motion (\ref{eomf4}) suggests
that the M2-brane charge can be supported by a non-trivial
harmonic 4-form in the 8-dimensional transverse space.  This
motivates us to make the following generalisation of the usual
M2-brane Ansatz:
\bea
d\hat s_{11}^2 &=& H^{-2/3}\, dx^\mu\, dx^\nu\, \eta_{\mu\nu} + H^{1/3}\,
ds_8^2\,,\nn\\
F_\4 &=& d^3 x\, \wedge dH^{-1} + m\, L_\4\,,
\eea
where $L_\4$ is an harmonic 4-form in the transverse 8-manifold $M_8$,
which has a  Ricci-flat metric $ds_8^2$.  Clearly
the Bianchi identity $dF_\4$ is trivially satisfied, since we have
$dL_\4=0$.  The equation of motion (\ref{eomf4}) implies
\be
\square H = -\ft1{48}m^2 \, L_\4^2\,,\qquad
L_4 = {* L_\4}\,.\label{m2source}
\ee
where $*$ is the Hodge dual with respect to $ds_8^2$, and one can then
easily verify that the Einstein equation is also
satisfied, provided that (\ref{m2source}) is
satisfied.\footnote{M2-brane solutions with a non-normalisable self-dual
4-form in the usual flat transverse 8-space were
constructed in \cite{deklm}; these were not supersymmetric, and they
had naked singularities.}  Next, we shall consider an M2-brane on an
explicit example of a complete non-compact 8-manifold of Spin(7)
holonomy that has a non-trivial normalisable self-dual harmonic
4-form.  We shall see that we can obtain a non-singular M2-brane solution.

\subsection{M2-brane on 8-manifold of Spin(7) holonomy}

    There are many examples of Ricci-flat 8-manifolds, including those
with hyper-K\"ahler and K\"ahler metrics.  Another possibility is to
consider 8-manifolds with Spin(7) holonomy; this is one of the
exceptional cases included in Berger's classification
\cite{berger}. There is a simple construction for one such example of a
complete Ricci-flat 8-manifold with Spin(7) holonomy
\cite{brysal,gibpagpop}.  The metric has cohomogeneity one, with the
level surfaces being 7-spheres described as a principal $SU(2)$ bundle
over $S^4$:
\be
ds_8^2 = \a^2\, dr^2 + \beta^2\, (\sigma_i - A^i)^2 + \gamma^2\,
d\Omega_4^2\,,\label{8met}
\ee
where $\a$, $\beta$ and $\gamma$ are functions of $r$, and as usual
$\sigma_i$ are left-invariant 1-forms of $SU(2)$.  The $SU(2)$
Yang-Mills potentials $A^i$ describe the BPST instanton on the unit
4-sphere whose metric is $d\Omega_4^2$.  One finds that the metric is
Ricci flat if \cite{brysal,gibpagpop}
\be
\a^2 = \Big(1 - \fft{a^{10/3}}{r^{10/3}}\Big)^{-1}\,,\qquad
\beta^2 = \ft{9}{100} r^2\,  \Big(1 - \fft{a^{10/3}}{r^{10/3}}\Big)\,,
\qquad \gamma^2 = \ft9{20} r^2\,.
\ee
The radial coordinate runs from $r=a$ to $r=\infty$.  At $r=a$, the
metric smoothly approaches $R^4\times S^4$, whilst at large $r$ the
level surfaces tend to the homogeneous squashed Einstein metric on
$S^7$.   Topologically, the manifold is an $R^4$ bundle over $S^4$.

    In order to look for a self-dual harmonic 4-form, it is useful to
introduce the quantities
\be
\ep_\3 \equiv\nu_1\wedge\nu_2\wedge\nu_3\,,\qquad
X_\3 \equiv \nu_i\wedge F^i\,,\qquad Y_\4 \equiv \ft12 \ep_{ijk}\,
\nu_i\wedge \nu_j\wedge F^k\,,
\ee
where $\nu_i\equiv \sigma_i-A^i$, and $F^i\equiv dA^i + \ft12
\ep_{ijk}\, A^j\wedge A^k$ is the Yang-Mills field strength for
the BPST instanton.  We have that $F^i$ is self-dual in the unit
4-sphere metric $d\Omega_4^2$, and $F^i\wedge F^i=6\Omega_\4$,
where $\Omega_\4$ is the volume form of the unit 4-sphere.  In
terms of these, we make the following Ansatz for a 3-form
potential $B_\3$ from which we shall seek to construct a
self-dual or anti-self-dual harmonic 4-form $G_\4=dB_\3$:
\be
B_\3 = f\, \ep_\3 + g\, X_\3\,,
\ee
where $f$ and $g$ are functions only of $r$. Thus we find
\be
G_\4 = f'\, dr\wedge \ep_\3 - (f+g)\, Y_\4 + g'\, dr\wedge X_\3 - 6g\,
\Omega_\4\,.
\ee
Hodge dualisation in the metric (\ref{8met}) gives
\be
{*(dr\wedge \ep_\3)} = \fft{\gamma^4}{\a\, \beta^3}\,
\Omega_\4\,,\qquad {\td Y_\4} = \fft{\a}{\beta}\, dr\wedge X_\3\,.
\ee
From this, it follows that imposing the duality condition ${\td G_\4}
= \eta\, G_\4$, where $\eta=\pm1$, gives the equations
\be
f' = -6 \eta\, g\, \a\, \b^3\, \gamma^{-4}\,,\qquad
g' = -\eta\, \a\, \beta^{-1}\, (f+g)\,,\label{fgeqs}
\ee
where $\eta=\pm1$ corresponds respectively to self-duality and
anti-self-duality.  Of course since $G_\4$ by construction is closed,
it follows that after imposing (anti) self duality, it will be harmonic.

    Defining $z=(a/r)^{10/3}$, these equations can be solved to give
\bea
\hbox{self-dual}:&& g = \fft{c_1\, z^{-1/5}}{1-z} +
                \fft{c_2\, z^{6/5}}{1-z}\,,\qquad
  f = -\ft65 c_1\, z^{-1/5} + \ft15 c_2\, z^{6/5} \,,
\label{4formharmonics}\\
\hbox{anti-self-dual}:&& g = c_1\, z^{-6/5} + c_2\, z^{1/5}\,,\qquad
   f = \ft15 c_1\, (1-6z)\, z^{-6/5} + \ft15 c_2\,
(z-6)\, z^{1/5}\,.\nn
\eea

   For suitable choices of the constants, it can be arranged that the
self-dual harmonic 4-form $G_\4$ has a non-diverging magnitude at
$r=a$, but it then does not fall off fast enough at $r=\infty$ to be
square integrable.

   On the other hand, the anti-self-dual harmonic 4-form $G_\4^-$
has magnitude given by the simple expression
\be
(G_\4^-)^2 = \fft{71680000 c_1^2}{243 a^8} +
   \fft{35840000 a^{4/3}\, c_2^2}{729 r^{28/3}}\,,
\ee
which is non-diverging at $r=a$ for all choices of its $c_1$ and $c_2$
integration constants.  In fact the case $c_2=0$ just corresponds to the
covariantly-constant harmonic 4-form that characterises a manifold of
Spin(7) holonomy; it can be expressed as $G_{abcd} = \bar\eta\,
\Gamma_{abcd}\, \eta$ where $\eta$ is the covariantly-constant spinor
(see \cite{gibpagpop} for further details).  Accordingly, the case of
greater interest to us is when $c_1=0$ and $c_2=1$, so that we get a
normalisable harmonic 4-form.  Expressed back in terms of $r$, this
solution is given by
\be
f = \ft15 \Big(\fft{a}{r}\Big)^{2/3}\, \Big( \fft{a^{10/3}}{r^{10/3}}
- 6\Big)\,,\qquad g =  \Big(\fft{a}{r}\Big)^{2/3}
\ee
From this, it follows that
\be
(G_\4^-)^2 = \ft{35840000 a^{4/3}}{729 r^{28/3}}\,.
\ee
It is easily seen that this is square-integrable.  It is topologically
non-trivial, since the expression for $B_\3$ becomes singular at
$r=a$.  (This can be seen from the fact that the coefficients of
$\ep_\3$ and $X_\3$ fail to vanish at $r=a$.)  In other words, it
cannot be written globally as the exterior derivative of a 3-form, and
so it is closed but not exact.

   If we take $L_4=G_\4^-$ in (\ref{m2source}), we  get the first integral
\be
\fft{\beta^3\, \gamma^4}{\a}\, H' = b + \fft{3 a^{4/3}\, m^2\, (7
r^{10/3} - 2 a^{10/3})}{5 r^{14/3}}\,,\label{xxxx}
\ee
where $b$ is a constant.  (Again we refer to footnote 4 for a
discussion of the necessary orientation reversal.)  The remaining
integration can be performed explicitly, giving an expression for $H$
in terms of elementary functions.  For generic values of the constant
$b$, the function $H$ diverges like $1/(r-a)$ as $r$ approaches $a$,
but this can be eliminated by choosing $b=-3m^2$.  After doing this,
we find that $H$ is regular everywhere in the interval $a\le
r\le\infty$, and it is given by
\bea
H &=& c -\fft{40000 m^2}{729 a^{16/3}\, r^{2/3} } \Big[9 -
\Big(\fft{a}{r}\Big)^{10/3} +
\fft{3\left( 1 -\fft{a^2}{r^2}\right)}{1- \left(\fft{a}{r}\right)^{10/3}}
\Big]\nn\\
&&+ \fft{32000 \sqrt{2\sqrt5}\, m^2}{243 a^6}\, \Big[
(\sqrt5-1)\, \arctan\left(\fft{\sqrt5+1+ 4\left(\fft{a}{r}\right)^{10/3}}{
\sqrt{2 \sqrt5(\sqrt5-1)}}\right)\nn\\
&&\qquad\qquad\qquad\qquad +
(\sqrt5+1)\, \arctan\left(\fft{\sqrt5-1+ 4\left(\fft{a}{r}\right)^{10/3}}{
\sqrt{2 \sqrt5(\sqrt5+1)}}\right) \Big]\,.
\eea
At large $r$, $H$ has the following behaviour,
\be
H = c + \fft{2\,10^5\, m^2}{3^7\, r^6} -
\fft{28\,10^4\, a^{4/3}\, m^2}{2673\, r^{22/3}} + \cdots\,.
\ee
It should be emphasised that the Chern-Simons flux term plays a crucial rule
for obtaining this regular M2-brane solution; the singularity would become
unavoidable if we were to set $m=0$ in (\ref{xxxx}).

    We may also consider the solution for an M2-brane supported by the
self-dual solution for $G_\4$, given by (\ref{4formharmonics}).  If we
choose the constant $c_1=0$, then we have the following
\be
r\rightarrow a : \qquad (G_\4^+)^2 \sim \fft{\rm const}{(r-a)^4}\,,\qquad
r\rightarrow \infty :\qquad (G_\4^+)^2 \sim \fft{\rm const}{r^{16}}\,,
\ee
and so it is normalisable at $r\rightarrow \infty$, but the integral
$\int \sqrt{g}\, d^8y\, (G_\4^+)^2$ diverges at
$r\rightarrow a$.  The solution for the function
$H$ has the following asymptotic behaviour:
\bea
r\rightarrow a: &&H\sim -\fft{\rm const}{(r-a)^3} \rightarrow -\infty\,,\nn\\
r\rightarrow \infty: && H \sim  1 + \fft{Q}{r^6} + \fft{\rm const}{r^{14}}
\,.
\eea
The solution is well-behaved for $r\rightarrow \infty$, with a
well-defined ADM mass, but it has a naked singularity when $r$
approaches $a$.  In order to avoid such a naked singularity, we can
instead choose the constants so that $c_1=-c_2$, in which case the
harmonic 4-form has the following asymptotic behaviour:
\be
r\rightarrow a :\qquad  (G_\4^+)^2 \sim {\rm const.}\,,\qquad
r\rightarrow \infty : \qquad(G_\4^+)^2 \sim \fft{\rm const}{r^{20/3}}\,.
\ee
This is normalisable in the region $r\rightarrow a$, but
non-normalisable for $r\rightarrow \infty$.  The function $H$ is given
by
\bea
H &=& c+ \fft{1600 m^2\, c_2^2\, y^2}{729 a^6\, (1+ y + y^2 y^3 + y^4)^3}
\Big(42 + 126 y + 231 y^2 + 357 y^3 + 504 y^4 +633 y^5 \nn\\
&&+744 y^6 + 809 y^7 + 828 y^8 + 801 y^9
+ 700 y^{10} + 525 y^{11} + 375 y^{12} + 250 y^{13} 
+ 150 y^{14} \nn\\
&&+ 75 y^{15} + 
25 y^{16}\Big)
-\fft{44800m^2\, c_2^2}{243a^6}\,\sum_{i=i}^4
\fft{y_i\,\log(y-y_i) + y_i^2\,\log(y-y_i)}{1+2y_i + 3y_i^2 + 4y_i^3}
\,,
\eea
where $y=(a/r)^{2/3}$ and $y_i$ are the four roots of the polynomial 
$1+y+y^2+y^3+y^4=0$.
It is easy to verify then that $H$ becomes a constant at $r=a$, and
behaves like $H\sim c + {\rm const}/r^{14/3}+ Q/r^6$ at large $r$. Thus
we see that $H$ does not fall off fast enough to give a well-defined ADM
mass in this case, which is a consequence of the fact that $G_\4$ is not
normalisable asymptotically.  However, the solution is regular
everywhere.

\section{D2-branes on 7-manifolds}

\subsection{General discussion}

      The D2-brane is supported by the 3-form potential in type IIA
theory.  It has a 7-dimensional transverse space.  At first sight, one
might think that the D2-brane is nothing but the vertical dimensional
reduction of the M2-brane we discussed above.  However, we can have a
non-trivial 7-dimensional space that is not merely the $S^1$ reduction
of one of the 8-dimensional spaces of the kind discussed in the
previous section, and so new kinds of deformed solution are possible
here.  The bosonic Lagrangian for type IIA supergravity is given by
\bea
{\cal L}_{\rm IIA} &=& \hat R\,{\hat *\oneone} -
\ft12 {\hat *d\phi}\wedge d\phi -\ft12
e^{-\phi}\, {\hat *F_\3\wedge F_\3} -\ft12 e^{\ft12\phi}\,
{\hat *F_\4\wedge F_\4} -\ft12 e^{\ft32\phi}\,
{\hat *F_\2\wedge F_2}\nn\\
&&+ \ft12 dA_\3\wedge dA_\3 \wedge A_\2\,,
\eea
where
\be
F_4 = dA_3 + A_\2\wedge dA_\1\,,\qquad
F_\3 = dA_\2\,,\qquad F_\2=dA_\1\,.
\ee
We shall look for a D2-brane solution for which $A_\1$ vanishes, and
hence we shall have $dF_\4=0$.  The equation of motion for $A_\3$ is
given by
\be
d (e^{\ft12\phi} \, {\hat * F_\4}) = F_\4\wedge F_\3
\ee
Thus if we can have a non-trivial harmonic 3-form in the 7-dimensional
transverse space, we can arrange to have a non-vanishing Chern-Simons term,
supporting the D2-brane.

      Let us consider the following D2-brane Ansatz
\bea
d\hat s_{10}^2 &=& H^{-5/8}\, dx^\mu\, dx^\nu\, \eta_{\mu\nu} +
H^{3/8}\, ds_7^2\,,\nn\\
F_4 &=& d^3x\wedge dH^{-1} + m_1\, {*L_\3}\,,\qquad
F_3 = m_2\, L_\3\,,\qquad \phi = \ft14 \log H\,,
\label{d2braneansatz}
\eea
where $L_\3$ is an harmonic 3-form in the Ricci-flat 7-metric $ds_7^2$.
It is straightforward to verify that all the type IIA equations of
motion are then satisfied provided that $m_1=-m_2\equiv m$, and that
\be
\square H = -\ft16 m^2\, L_\3^2\,.\label{d2heq}
\ee

\subsection{D2-brane on 7-manifold with $G_2$ holonomy}

   We may consider a simple example of a complete non-compact
7-manifold with a Ricci-flat metric of $G_2$ holonomy.  It is given by
\cite{brysal,gibpagpop}
\be
ds_7^2 = \a^2\, dr^2 + \beta^2\, (\sigma_i - \ft12 \Sigma_i)^2 +
\gamma^2\, \Sigma_i^2\,,
\ee
where the functions $\a$, $\beta$ and $\gamma$ are given by
\be
\a^2 = \Big(1-\fft{a^3}{r^3}\Big)^{-1}\,,\qquad
\beta^2 = \ft19 r^2\,  \Big(1-\fft{a^3}{r^3}\Big)\,,\qquad
\gamma^2 = \ft1{12} r^2\,.
\ee
Here $\Sigma_i$ and $\sigma_i$ are two sets of left-invariant 1-forms
on two independent $SU(2)$ group manifolds.  The level surfaces
$r=$constant are therefore $S^3$ bundles over $S^3$.  This bundle is
trivial, and so in fact the level surfaces are topologically
$S^3\times S^3$.  The radial coordinate runs from $r=a$ to $r=\infty$.

    If we define an orthonormal frame by
\be
e^0= \a\, dr\,,\qquad e^i =\gamma\, \Sigma_i\,,\qquad e^{i+3} = \beta\,
\nu_i
\ee
where $\nu_i\equiv \sigma_i -\ft12 \Sigma_i$, then one can read off from
results in \cite{gibpagpop} that there is a single covariantly-constant
spinor, which satisfies the projection conditions
\be
(\Gamma_{04}-\Gamma_{23})\, \eta= (\Gamma_{05}-\Gamma_{31})\, \eta=
(\Gamma_{06}-\Gamma_{12})\, \eta= 0\,.
\ee
As discussed in \cite{gibpagpop}, one can then construct a
covariantly-constant 3-form $Q_\3$, defined by $Q_{abc}=\bar\eta\,
\Gamma_{abc}\, \eta$.  This turns out to be
\be
Q_\3 = e^0\wedge e^i\wedge e^{\td i} + \ft12 \ep_{ijk}\, e^i\wedge
e^{\td j}\wedge e^{\td k} - e^1\wedge e^2\wedge e^3\,,
\ee
where we have defined $e^{\td i} \equiv e^{i+3} = \beta \, \nu_i$ for
$i=1,2,3$.

    The form of $Q_\3$ suggests a natural Ansatz for trying to find
further harmonic 3-forms.  Thus we let
\be
G_\3 = f\, dr\wedge \nu_i\wedge \Sigma_i + \ft12 g\, \ep_{ijk}\,
\nu_i\wedge \nu_j\wedge \Sigma_k + h\, \Sigma_1\wedge\Sigma_2\wedge
\Sigma_3\,,
\ee
where $f$, $g$ and $h$ are functions only of $r$.
The condition $dG_\3=0$ implies
\be
4h'- 3f=0\,,\qquad g'+f=0\,,
\ee
giving a first integral
\be
h= 3b- \ft34 g\,,
\ee
where $b$ is an arbitrary constant.  The condition $d{* G_\3}=0$
gives the equation
\be
\Big(\fft{f\, \beta\, \gamma}{\a}\Big)'   + \fft{g\, \a\,
\gamma}{\beta} - \fft{h\, \a\, \beta^3}{4\gamma^3} =0\,.
\ee
We find that the general solution is
\be
g=\fft{b\, (r^3-4a^3) + c_1/r + c_2\, r^3(4r^3- 7a^3)}{r^3-a^3}\,,
\ee
together with
\be
f=-g'\,,\qquad h= 3b- \ft34 g\,.
\ee

   The magnitude of $G_\3$ is given by
\be
G_\3^2 = 6\Big( \fft {3f^2}{\a^2\, \beta^2\, \gamma^2} + \fft{3
g^2}{\beta^4\, \gamma^2} + \fft{h^2}{\gamma^6}\Big)\,.
\ee
It follows that $G_\2^2$ will diverge at $r=a$ in general, but it will
be non-singular if the constants are chosen so that
\be
 c_1 = 3 a^4\, (b + a^3\, c_2)\,.\label{con1}
\ee
The original covariantly-constant 3-form $Q_\3$ is obtained if one
additionally sets $b=-a^3\, c_2$.  In this case we would find that
$G_\3^2$ was simply a constant.  Instead, we can get a harmonic 3-form
that falls off at large $r$ if we still impose (\ref{con1}), but now
additionally choose $c_2=0$.  This gives a harmonic 3-form with the
following asymptotic behaviours:
\be
G_\3^2 \sim \hbox{const} + (\hbox{const})\, (r-a) + \cdots
\ee
as $r$ approaches $a$, and
\be
G_\3^2 \sim \fft{\hbox{const}}{r^6} + O(1/r^9)
\ee
as $r$ approaches infinity.  This is almost, but not quite,
normalisable, for $r\rightarrow \infty$.  The function $H$ can be solved
explicitly, given by
\bea
H&=& c + \fft{108m^2\, b^2\, (a+r)}{a^3\, r^3\, 
(r^2 + a\, r + a^2)^3}\,
\Big(16r^7 + 24a\, r^6 + 48 a^2\, r^5
+ 47 a^3\, r^4\ + 54 a^4\, r^3\nn\\ 
&&+ 36a^5\, r^2 + 18 a^6\, r + 9a^7\Big)
+1152\sqrt3\, m^2\, b^2\, a^{-4}\, {\rm arctan} \fft{2r +a}{\sqrt3\,
a} \,.
\eea
Thus the function $H$ is perfectly non-singular for $r$ running from $a$
to infinity.  It approaches a positive constant for $r\rightarrow a$,
with the asymptotic behaviour
\be
H \sim {\rm const} + \fft{\rm const}{r^4} +
\fft{\rm const}{r^5}\,,
\ee
for $r\rightarrow \infty$.  Thus the function $H$ does not fall off fast
enough to have a well-defined ADM mass, which is a consequence of the
fact that $G_\3^2$ is linearly non-normalisable in the asymptotic
region.

\section{Further examples}

\subsection{Type IIA and type IIB strings on 8-manifolds}

    There are two different possibilities for obtaining deformed
string solutions in ten dimensions, depending upon whether we consider
type IIA strings or type IIB strings.  In each case, the
8-dimensional transverse space will first be replaced by a Ricci-flat
manifold $M_8$.

   In type IIA, the string can be obtained as a diagonal dimensional
reduction of the eleven-dimensional M2-brane.  Since this leaves the
transverse space intact, the deformed solution follows directly from
our results for the deformed M2-brane in section 5.  Thus the solution
is given by
\bea
d\hat s_{10}^2 &=&  H^{-3/4}\, (-dt^2 + dx^2) +
H^{1/4}\, ds_8^2\,,\nn\\
F_\3 &=& dt\wedge dx \wedge dH^{-1}\,,\qquad \phi = -\ft12 \log
H\,,\\
F_\4 &=& m\, L_\4\,,\qquad F_\2=0\,.\nn
\eea
This satisfies the equations of motion of type IIA supergravity
provided that $ds_8^2$ is a Ricci-flat on the transverse space $M_8$,
$L_\4$ is a self-dual harmonic 4-form on $M_8$, and $H$ satisfies
\be
\square \, H = -\ft1{48} m^2\, L_\4^2\,.
\ee
(See section 6 for the convention and the Lagrangian of type IIA
supergravity.)  All the features of the deformed M2-brane solutions will
carry over directly to these deformed type IIA string solutions.
The resulting type IIA string is then completely regular, as
in the case of M2-brane.

   The situation is quite different if we consider strings in the type
IIB theory instead.  Now, for an NS-NS string, the Ansatz will be
\bea
d\hat s_{10}^2 &=& H^{-3/4}\, (-dt^2 + dx^2) +
H^{1/4}\, ds_8^2\,,\nn\\
F_\3^\ns &=& dt\wedge dx \wedge dH^{-1}\,,\qquad \phi = -\ft12 \log
H\,,\\
F_\3^\rr &=& m\, L_\3\,,\qquad F_\5 = m\, {*L_\3} + m\, H^{-1}\,
dt\wedge dx\wedge L_\3\,,\nn
\eea
where $ds_8^2$ is a Ricci-flat 8-metric on a manifold $M_8$ and $L_\3$
is an harmonic 3-form on $M_8$.  The notation for the type IIB fields
is the same as in section 2, with $F_\3^\rr$ denoting the R-R 3-form,
and $F_\3^\ns$ denoting the NS-NS 3-form.

    Substituting into the type IIB equations of motion, (given in
section 2,) we find that the above Ansatz for a deformed NS-NS
string solve the type IIB equations of motion, provided that
$L_\3$ is harmonic and that $H$ satisfies
\be
\square H = -\ft16 m^2\, L_\3^3\,.
\ee
If there is a normalisable 3-form in $M_8$ then it will be possible to
construct a string solution with no singularities.  As usual, if the
manifold $M_8$ has a special holonomy, so that it admits
covariantly-constant spinors, then the string solution can still
preserve some supersymmetry.

\subsection{D4-branes on 5-manifolds}

    Another example that can be constructed is a deformed D4-brane
solution in the type IIA theory.  In this case the transverse
space is five-dimensional.  There are no irreducible manifolds
$M_5$ of special holonomy, but for completeness we may consider
this example too.  Thus we make the Ansatz
\bea
d\hat s_{10}^2 &=& H^{-3/8}\, dx^\mu\, dx^\nu\, \eta_{\mu\nu} +
    H^{5/8}\, ds_5^2\,,\nn\\
F_\4 &=& {* dH} \,,\qquad \phi = -\ft14 \log H\,,\\
F_\2 &=& m\, L_\2\,,\qquad F_\3 = m\, {*L_\2}\,.
\eea
Substituting into the equations of motion of type IIA supergravity,
(given in section 6,) we find that this gives a deformed D4-brane
solution provided that $L_\2$ is an harmonic 2-form in the manifold
$M_5$ with Ricci-flat metric $ds_5^2$, and that $H$ satisfies
\be
\square H = -\ft12 m^2\, L_\2^2\,.
\ee

   One possible choice for $M_5$ is to take the product $M_5=M_4\times
\R$, where $M_4$ is any Ricci-flat 4-manifold.  For example, we
can take $M_\4$ to be Eguchi-Hanson (Taub-NUT) metric, in which
case the self-dual  (anti-self-dual) harmonic 2-form given in
(\ref{l2exp}) will also be harmonic in $M_5$.  It will no longer
be normalisable, owing to the non-compact 5'th direction upon
which it does not depend.  One could take the 5'th direction to be
$S^1$ instead, in which case it would still be normalisable in
$M_5=M_4\times S^1$.

\section{Conclusions}

   In this paper we have introduced a general procedure for obtaining
families of deformations of certain of the standard $p$-brane
solutions in supergravity.  The method is applicable to cases where
the field strength that supports the standard brane solution has
Chern-Simons type terms (or transgressions) in its Bianchi identity or
equation of motion.  The deformed solution is obtained by first
replacing the flat transverse space by a manifold $M_n$ with a
Ricci-flat metric, and then using a harmonic form in $M_n$ to give a
non-vanishing flux, which introduces fractional branes, for the fields
appearing bilinearly in the Chern-Simons terms. The cases of principle
interest are where $M_n$ admits covariantly-constant spinors, since
means that the deformed solution may still preserve some
supersymmetry, thus providing gravity solutions that are dual to
super-Yang-Mills theories, possibly with less than maximal
supersymmetry.  Usually, one would also want to take $M_n$ to be a
complete non-compact manifold. If the relevant harmonic form is
normalisable (square integrable), then the deformed brane solution can
become completely free of singularities. These examples are of special
interest since now the supergravity description is valid in the whole
domain of space-time and thus can provide important information about
the properties of the dual field theory (in the infra-red regime),
such as confinement and chiral symmetry breaking.

    The method that we have developed here for obtaining the deformed
brane solutions is a generalisation of a procedure that has been much
discussed recently in the context of D3-branes, in which non-zero flux
for the R-R and NS-NS 3-forms (fractional branes) is turned on
\cite{klebtsey,klebstra,ganpol,zaytse}.  These examples are of special
interest, since they provide information on strongly coupled
four-dimensional $N=1$ super-Yang-Mills theory.   One of the
resolutions of the D3-brane is discussed in \cite{zaytse}. 
In this particular case the harmonic 3-form used in the deformed
solution is not normalisable, and so the naked singularity, of the
repulson-type \cite{repulson,jopepo}, in the D3-brane solution is not
eliminated.

    In our more general discussion for a variety of other dimensions, we
found several explicit examples where brane singularities could be
completely resolved, including the heterotic 5-brane, the dyonic string,
the M2-brane, the type IIA string and the D2-brane.  In general, such a 
complete resolution can be achieved whenever one has a normalisable 
harmonic form to supply the required flux.

    A particularly interesting example is that of heterotic 5-brane,
where the resolution is achieved by taking an Abelian $U(1)$ gauge field
to have a flux proportional to the normalisable self-dual (or
anti-self-dual) harmonic 2-form of the Eguchi-Hanson (or Taub-NUT)
metric.  Since the Eguchi-Hanson metrics can provide local resolutions
of the $T^4/Z_2$ orbifold singularities, these examples provide a
completely non-singular solution of the 5-brane on the K3 manifold, with
each of the sixteen 2-form field strengths in the Cartan subalgebra of
the heterotic string equal to the localised self-dual harmonic 2-form of
each Eguchi-Hanson instanton. The non-singular gravity solution may
provide a viable candidate for studying the dual six-dimensional field
theory with $N=2$ supersymmetry.  A preliminary analysis indicates that
the spectrum may not have a bound state, and that it is continuous with
a mass gap.

   The dyonic string is another example of a completely regular deformed
solution, whose three-dimensional field-theory dual exhibits a
bound-state spectrum and thus a confinement.  Interestingly, the example
of the tensionless string, which exhibits a repulson-type singularity
\cite{repulson}, can now be completely resolved, yielding a non-singular
solution in the whole region, thus providing an alternative to
Yang-Mills instanton resolution\cite{dulupo} and the \enhancon\
resolution \cite{jopepo}.

    There are a number of possible generalisations \cite{clp} of the
brane-resolution mechanism constructed in this paper, which may involve
more then one harmonic (normalisable) form as well as possible
intersecting $p$-brane configurations.  These would provide novel
regular supergravities as candidate duals for field theories with less
supersymmetry, and in particular, the $N=1$ supersymmetric
four-dimensional examples.

\section*{Acknowledgement} 

We are very grateful to Gary Gibbons for extensive discussions about
non-compact Ricci-flat metrics, and to Leopoldo Pando-Zayas for useful
discussion on D3-branes on conifolds. We also benefited from discussions
with Christian R\"omelsberger and Barthomeu Fiol.  We thank Rutgers High
Energy Theory Group (M.C.) and the organisers of the ``Semestres
Cordes'' at the Centre Emile Borel of the Institut Henri Poincar\'e
(C.N.P.) for support and hospitality during the course of this work.

\end{document}